\documentclass{moriond}
\usepackage{amsmath}
\usepackage{lineno}
\def\rd{{\rm{d}}}

\def\be{\begin{equation}}
\def\ee{\end{equation}}
\def\bea{\begin{eqnarray}}
\def\eea{\end{eqnarray}}

\newcommand{\Photo}{\includegraphics[height=35mm]{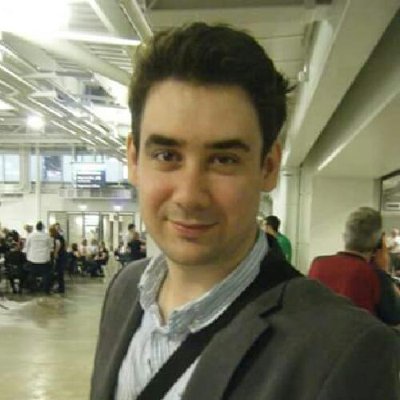}}

\begin{document}
\vspace*{4cm}
\title{{{INCLUSIVE SINGLE GAUGE BOSON PRODUCTION IN $\sqrt{s}=7,\,8$ AND $13$~TeV PROTON-PROTON COLLISIONS}}}

\author{{{W. BARTER$^*$}}\\on behalf of the ATLAS, CMS, and LHCb collaborations}

\address{$^*$European Organisation for Nuclear Research (CERN), Geneva, Switzerland\\$ $}

\maketitle\abstracts{
We report LHC measurements of single $W$ and $Z$ boson production at different centre-of-mass energies. Together, the ATLAS, CMS, and LHCb detectors cover an absolute rapidity range from 0 to 4.5, enabling precision tests and studies of Standard Model physics across different kinematic regions. We report total, fiducial and differential cross-section measurements and their ratios. These results allow tests of the performance of different parton distribution functions (PDFs), and can be used to further constrain the PDFs. In addition, the results are compared to theoretical predictions which use different approaches to model effects within quantum chromodynamics. We also report measurements of the forward-backward asymmetry in $Z$ boson decays, and measurements of angular coefficients, which provide sensitivity to the electroweak mixing angle.}

\section{Cross-section measurements}
Measurements of gauge boson production cross-sections at the LHC provide sensitivity to many different areas of particle physics. This article concentrates on measurements of inclusive $W$ and $Z$ boson production.\footnote{This article includes the contributions from virtual photons within the label $Z$.} Total, fiducial, and differential cross-sections as a function of the boson rapidity are well predicted within the Standard Model at the partonic level, allowing study of the internal structure of the proton and PDFs~\cite{pdf4lhc}, across a wide range of Bjorken-$x$. Differential cross-section measurements can also be used to study modelling of $pp$ collisions, allowing the tuning of different event generators used to simulate proton-proton collisions. Such studies will be crucial for any future LHC measurement of the $W$ boson mass.\cite{mw} Ratios of cross-section measurements at different $pp$ collision energies also provide sensitivity to potential physics beyond the Standard Model.\cite{mm}

\subsection{Selected measurements from LHC Run 2}
The ATLAS, CMS and LHCb collaborations have all made measurements of gauge boson production in $pp$ collisions at 13 TeV.\cite{at13,cms13,lhcb13} The ATLAS and CMS collaborations have presented measurements of $W$ and $Z$ boson production in fiducial regions corresponding to the acceptance of their detectors using decays to final states including electrons or muons. For ATLAS, the fiducial acceptance corresponds to the leptons having transverse momentum, $p_{\rm{T}}^{\ell} > 25$~GeV, and absolute pseudorapidity, $|\eta_{\ell}| < 2.5$. For $Z$ boson production the dilepton invariant mass $m_{\ell\ell}$ is required to be between $66<m_{\ell\ell}<116$~GeV, and for $W$ boson production the transverse mass, $m_{\rm{T}}$, is required to be above 50~GeV. At CMS, the fiducial requirements are different for electrons and muons, and are optimised to the CMS detector acceptance. ATLAS and CMS have also measured total cross-sections where the fiducial measurements are extrapolated to the full phase space using simulation. The requirement on the di-lepton invariant mass is retained since the differential cross-section increases significantly at low dilepton invariant masses due to the virtual photon contribution. At CMS this corresponds to a window between  60 and 120 GeV. The LHCb collaboration have made measurements of Z boson production using muons in a fiducial region corresponding to the forward region occupied by the LHCb detector: muons are required to have $2.0 < \eta^\mu < 4.5$ and $p_{\rm{T}}^\mu > 20$~GeV, and the invariant mass of the dimuon system is required to be between 60 and 120~GeV.

\begin{figure}
\begin{center}
  \includegraphics[width=0.95\textwidth]{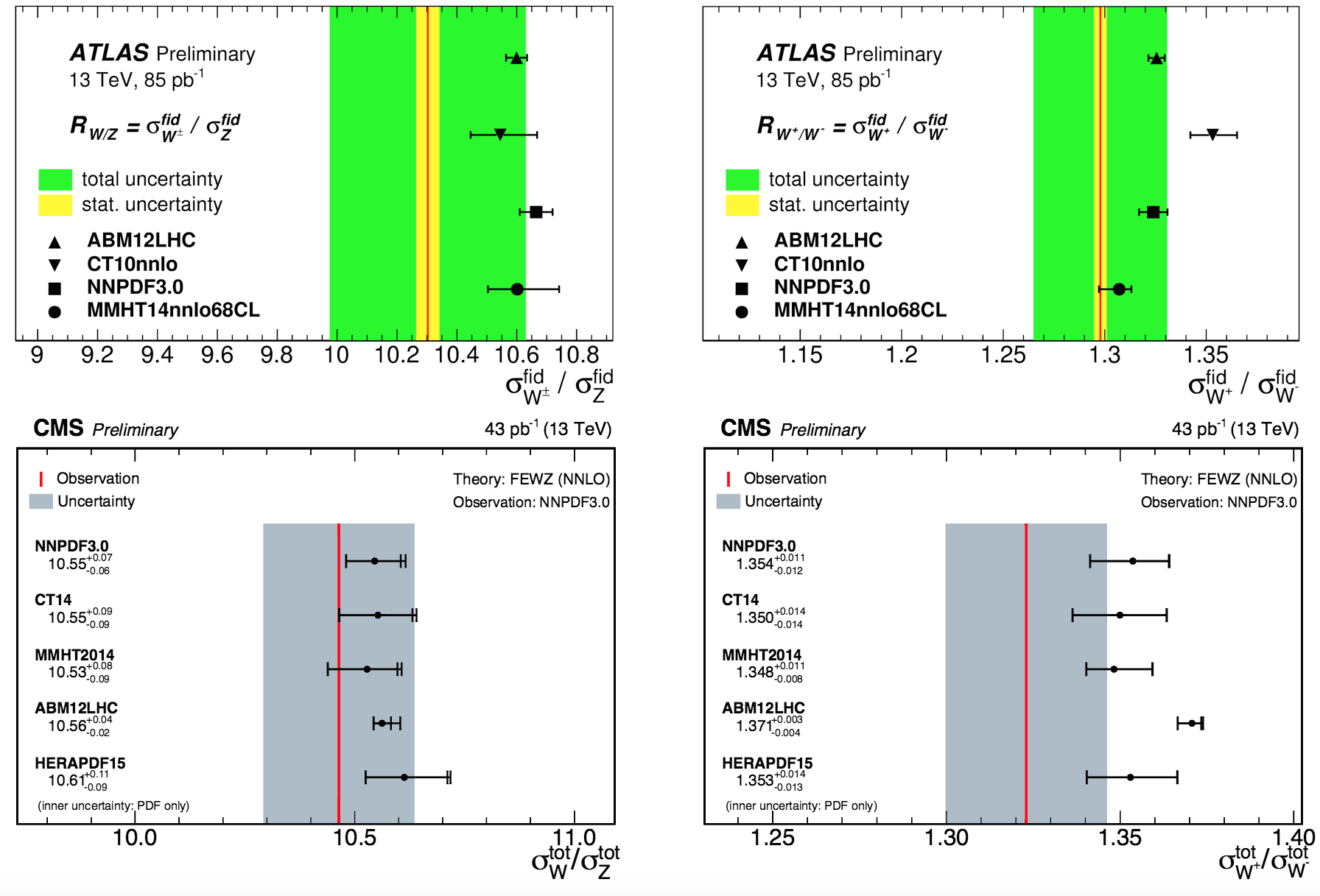}
\caption{(Top) measurements of fiducial cross-sections ratios by the ATLAS collaboration, and (bottom) measurements of total cross-section ratios by the CMS collaboration. The data are shown as bands, with the inner band corresponding to the statistical uncertainty and the outer band corresponding to the total uncertainty. NNLO predictions calculated using different PDF sets are shown as points. From the original articles~$^{4,\,5}$, where more detail is given about the theoretical predictions compared to the data.\label{fig:13_rat}}
\end{center}
\end{figure}

\begin{figure}
\begin{center}
  \includegraphics[width=0.75\textwidth]{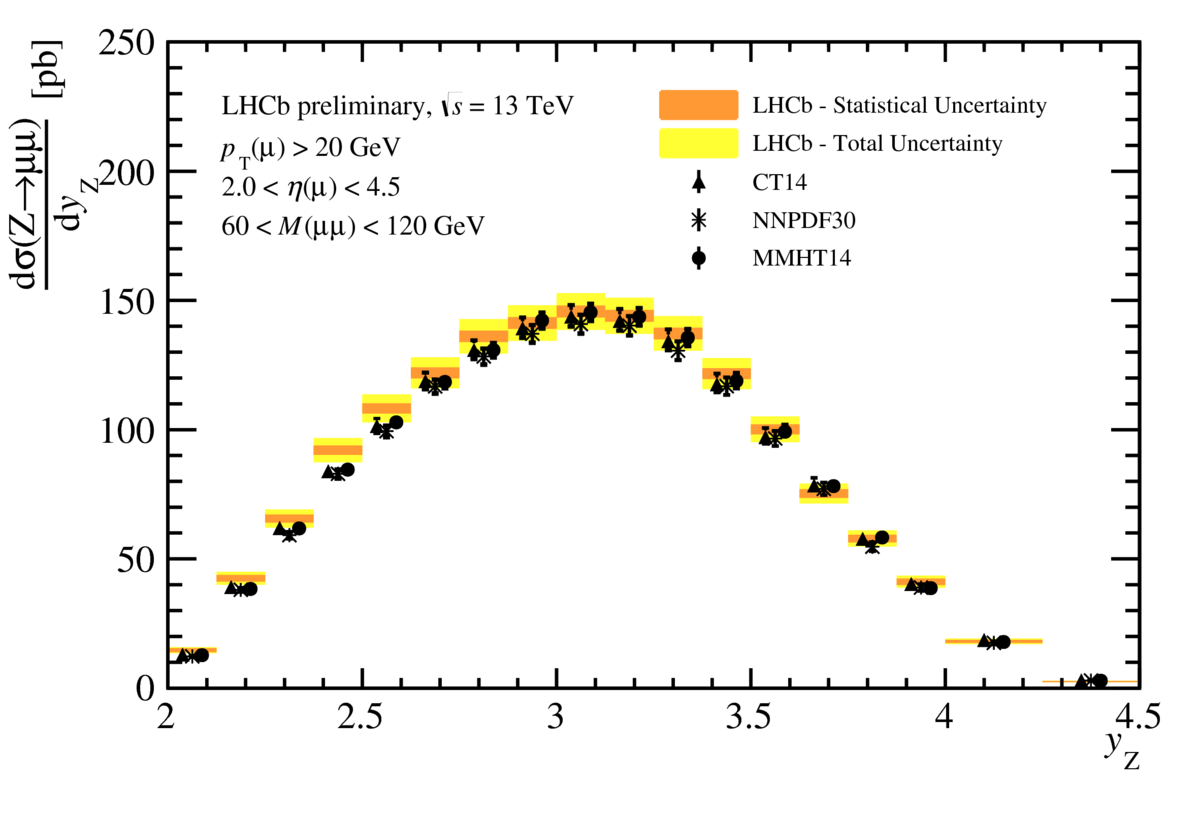}
\caption{Measurements of the differential cross-section for $Z$ boson production as a function of the boson rapidity. The data are shown as bands, with the inner band corresponding to the statistical uncertainty and the outer band corresponding to the total uncertainty. NNLO predictions calculated using different PDF sets are shown as points. From the original article~$^6$, where more detail is given about the theoretical predictions compared to the data.\label{fig:lhcb_rap13}}
\end{center}
\end{figure}

All the LHC experiments have compared the measurements to theoretical predictions calculated at next-to-next-to leading order (NNLO) in perturbative quantum chromodynamics (pQCD) based on different PDF sets and found good agreement. Cross-section ratios between $W$ and $Z$ boson production, and between $W^+$ and $W^-$ production, are shown for ATLAS and CMS in Figure~\ref{fig:13_rat}. The differential cross-section for $Z$ boson production as a function of the boson rapidity as measured at LHCb is shown in Figure~\ref{fig:lhcb_rap13}. With increased precision future measurements from all three experiments will provide important constraints on PDFs.

\subsection{Selected measurements from LHC Run 1}
There are many important measurements of $W$ and $Z$ boson production produced by the LHC experiments for both $\sqrt{s} = 7$ and $8$~TeV $pp$ collisions. This article selects a few results of interest from recent publications from each of the experiments.

\begin{figure}
\begin{center}
  \includegraphics[width=0.495\textwidth]{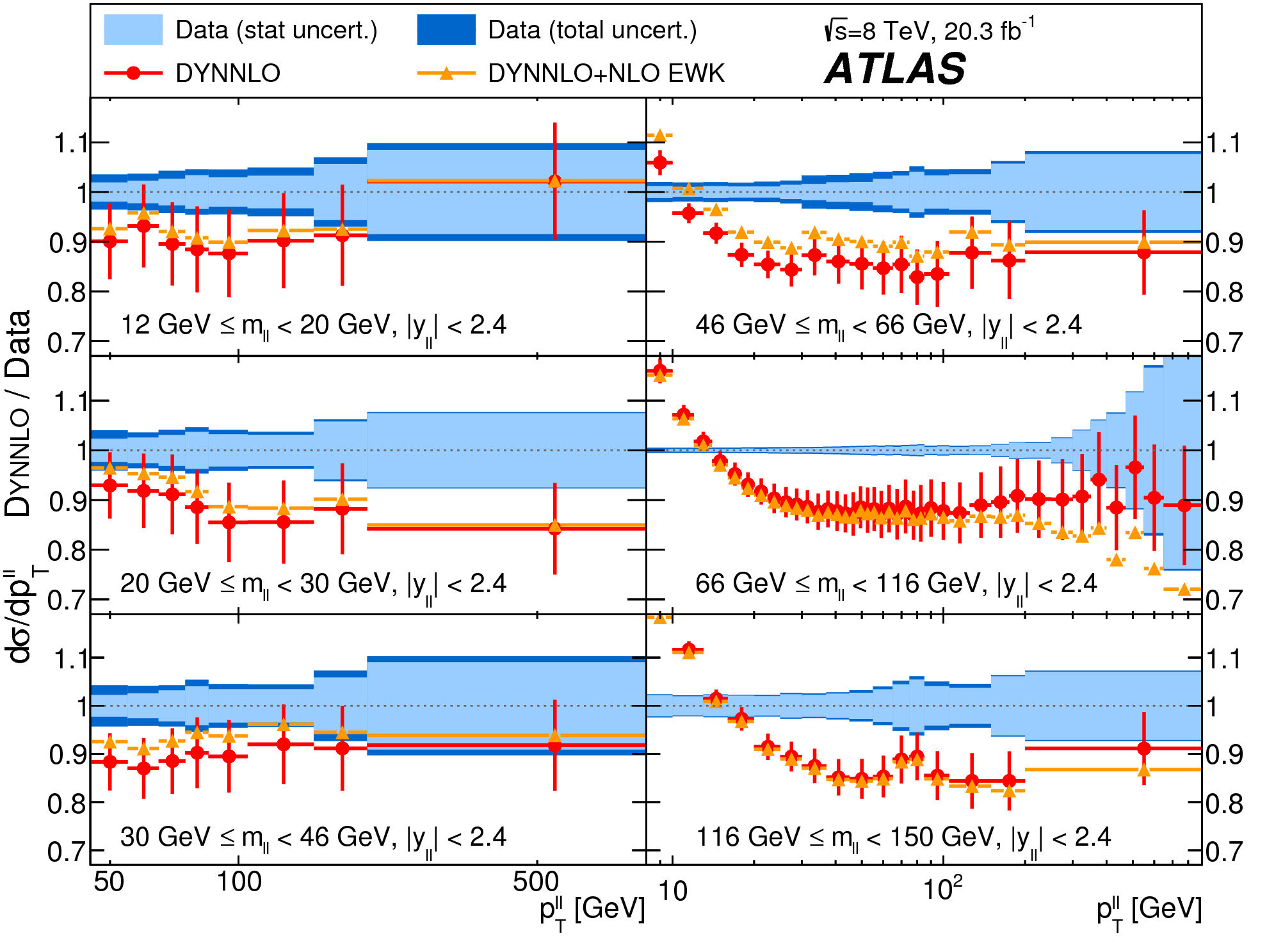}
  \includegraphics[width=0.495\textwidth]{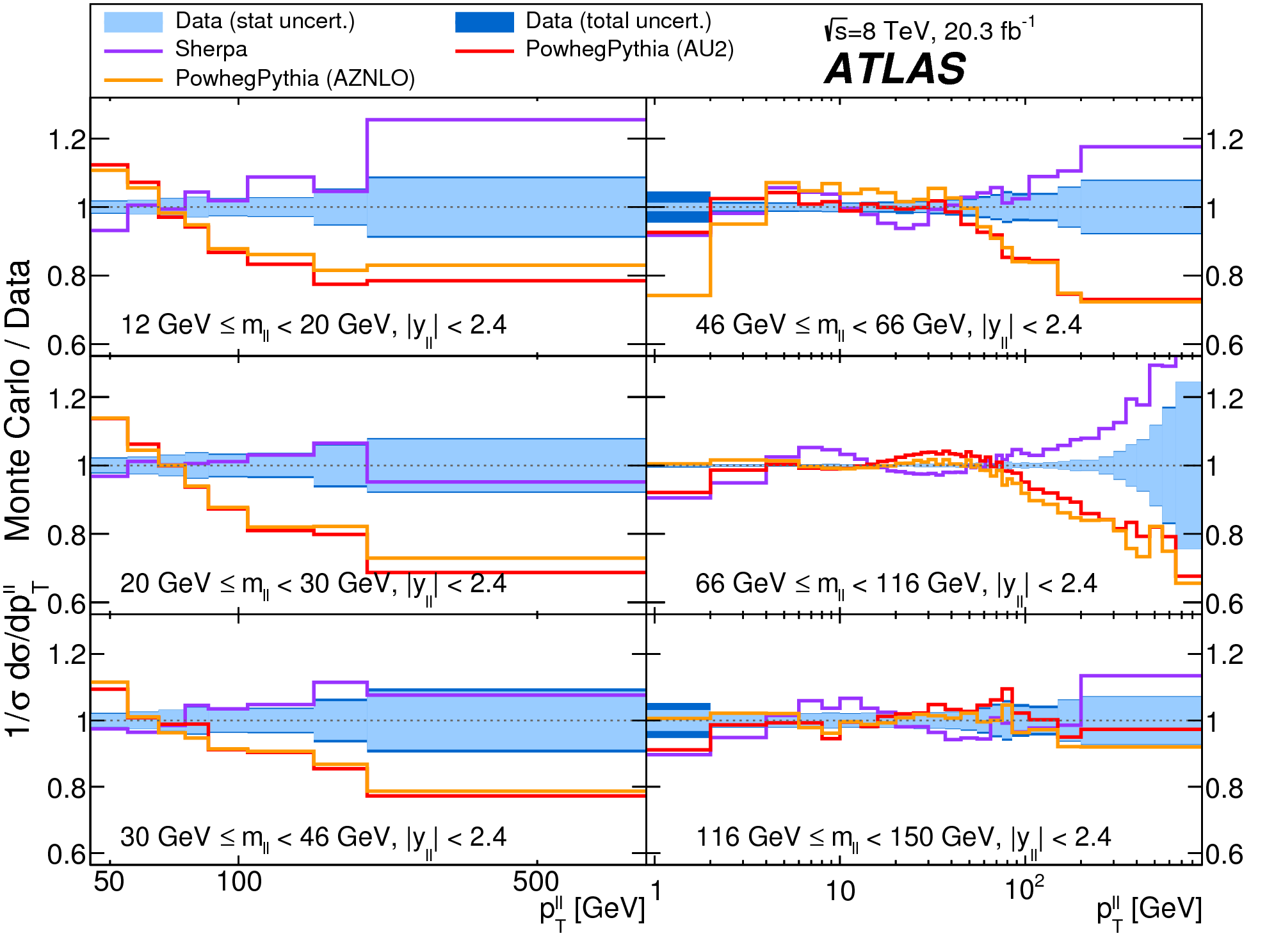}
\caption{Measurements of the differential cross-section for Drell-Yan production as a function of the dilepton transverse momentum, in different invariant mass bins. The data are shown as bands, with the inner band corresponding to the statistical uncertainty and the outer band corresponding to the total uncertainty. The left hand plot compares the data to fixed order predictions, whereas the right hand plot compares the data to different event generators. From the original article~$^8$, where more detail is given about the theoretical predictions compared to the data.\label{fig:atlasdy}}
\end{center}
\end{figure}

The ATLAS experiment has reported the measurement of the $Z$ boson transverse momentum and $\phi_\eta^*$ distributions~\cite{phist}$^,\,$\footnote{The $\phi_\eta^*$ variable is strongly correlated to the boson transverse momentum but is less effected by experimental resolution as it depends only on angular variables. Here only transverse momentum results are shown.} in $8$~TeV $pp$ collisions.\cite{at8} The measurement has been performed differentially in the dilepton invariant mass and rapidity. The precise fiducial definitions are given in the original article~\cite{at8}. The results are compared to theoretical predictions generated using different Monte Carlo event generators, and fixed order predictions (at NNLO in pQCD) calculated using DYNNLO~\cite{dynnlo}. Selected comparisons are shown in Figure~\ref{fig:atlasdy}. Even at high transverse momentum, significant discrepancies are seen between the fixed order predictions and the data. This disagreement appears to be reasonably independent of the dilepton rapidity and mass and transverse momentum for $p_{\rm{T}}^{\ell\ell} > 50$~GeV, with theory consistently $\sim 10\%$ lower than the measured differential cross-section. The event generator predictions agree best with data when the dilepton invariant mass is close to the $Z$ boson mass ($m_{Z}$), and the best agreement is seen for $p_{\rm{T}}^{\ell\ell} < m_{Z}$. Since the greatest sensitivity to the $W$ boson mass, $m_W$, comes from events with small boson transverse momentum, such behaviour is encouraging for future $W$ boson mass measurements.

\begin{figure}
\begin{center}
  \includegraphics[width=0.495\linewidth]{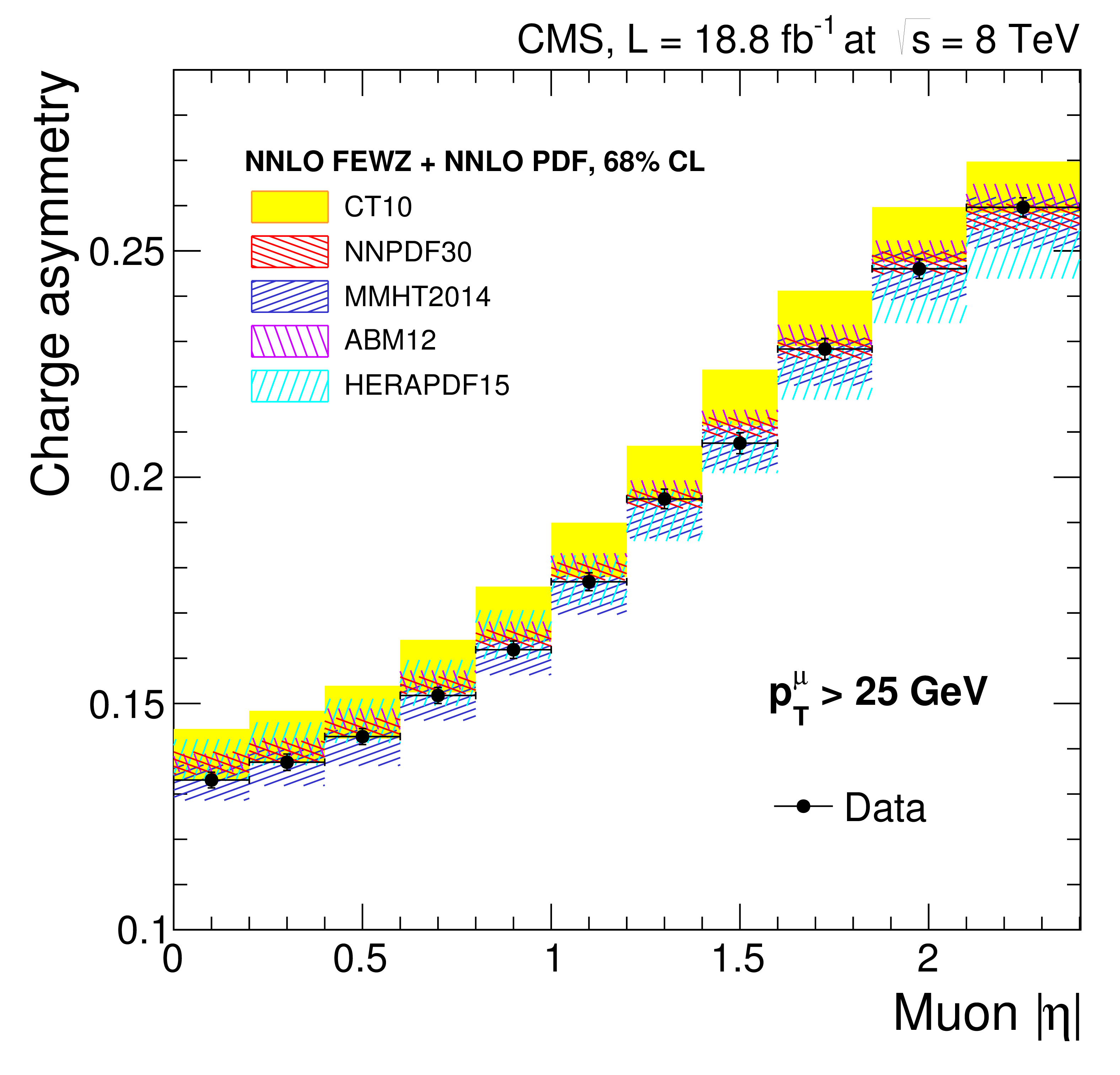}
  \includegraphics[width=0.495\linewidth]{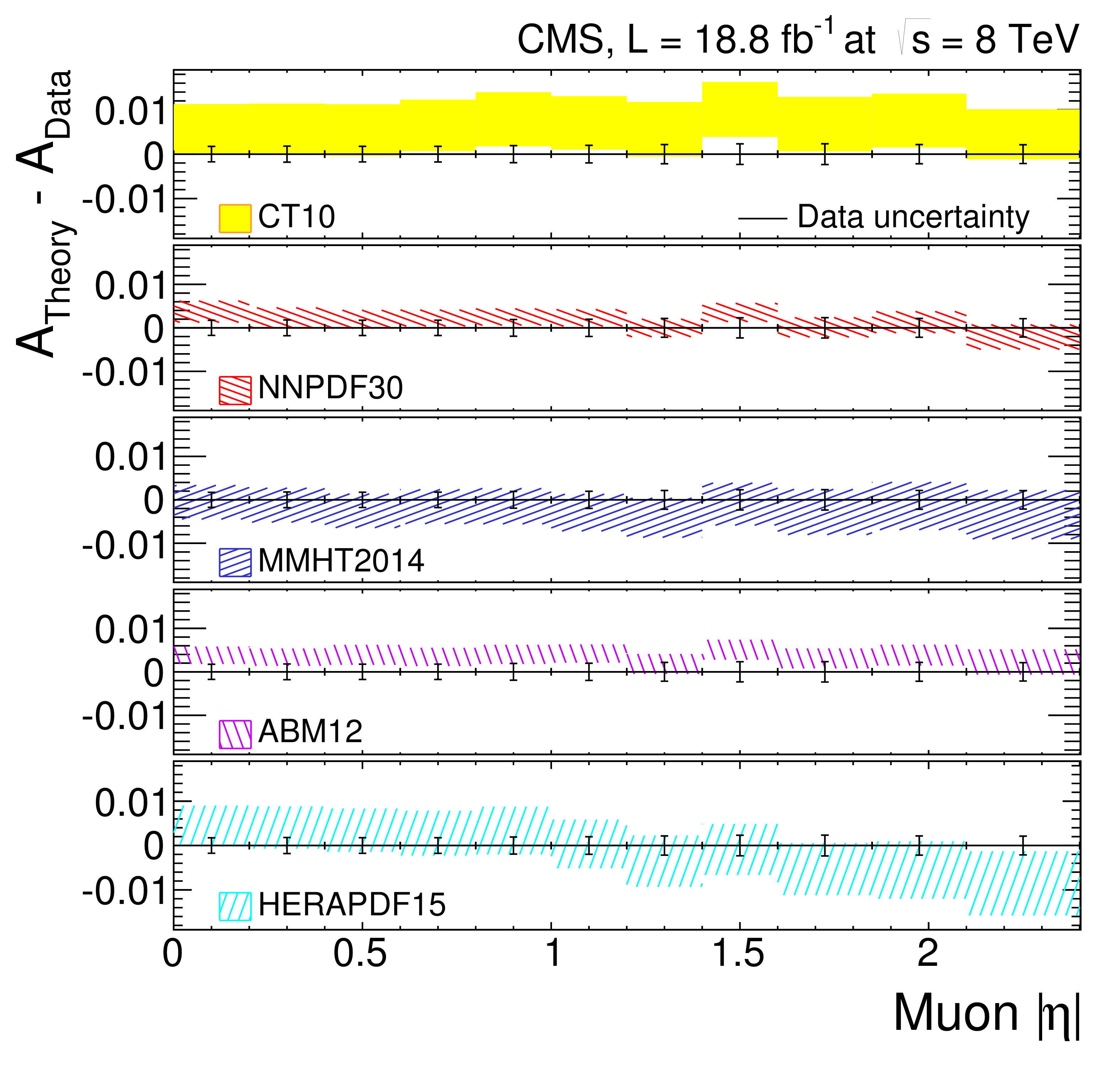}
\caption{Measurements of the lepton charge asymmetry measured by the CMS experiment. The points correspond to the measured result, while the shaded bands correspond to different predictions at NNLO using different PDF sets. From the original article~$^{10}$, where more detail is given about the theoretical predictions compared to the data.\label{fig:cms_asym}}
\end{center}
\end{figure}

The CMS experiment has reported a measurement of inclusive $W$ boson production in $8$~TeV $pp$ collisions, using events where the $W$ boson decays to a muon and a neutrino.\cite{cms8} The measurement has been performed for a fiducial cross-section defined by $|\eta^\mu|<2.4$ and $p_{\rm{T}}^\mu > 25$~GeV. The measurement is made differentially in bins of the muon pseudorapidity separately for both charges, allowing the differential lepton charge asymmetry, $A \equiv ({\rm{d}}\sigma^+ - {\rm{d}}\sigma^-) / ({\rm{d}}\sigma^+ + {\rm{d}}\sigma^-)$ to be measured in addition. This variable is particularly sensitive to the modelling of the ratio of the PDFs for the up and down valence quarks. The measurements are compared to theoretical predictions calculated at NNLO using FEWZ~\cite{fewz} for different PDF sets, and show good agreement; the asymmetry is shown in Figure~\ref{fig:cms_asym}. However, the uncertainties on the data are smaller than the uncertainties on the theoretical predictions, indicating that this measurement will play an important role providing constraints in future PDF fits.

\begin{figure}
\begin{center}
\includegraphics[width=0.5\linewidth]{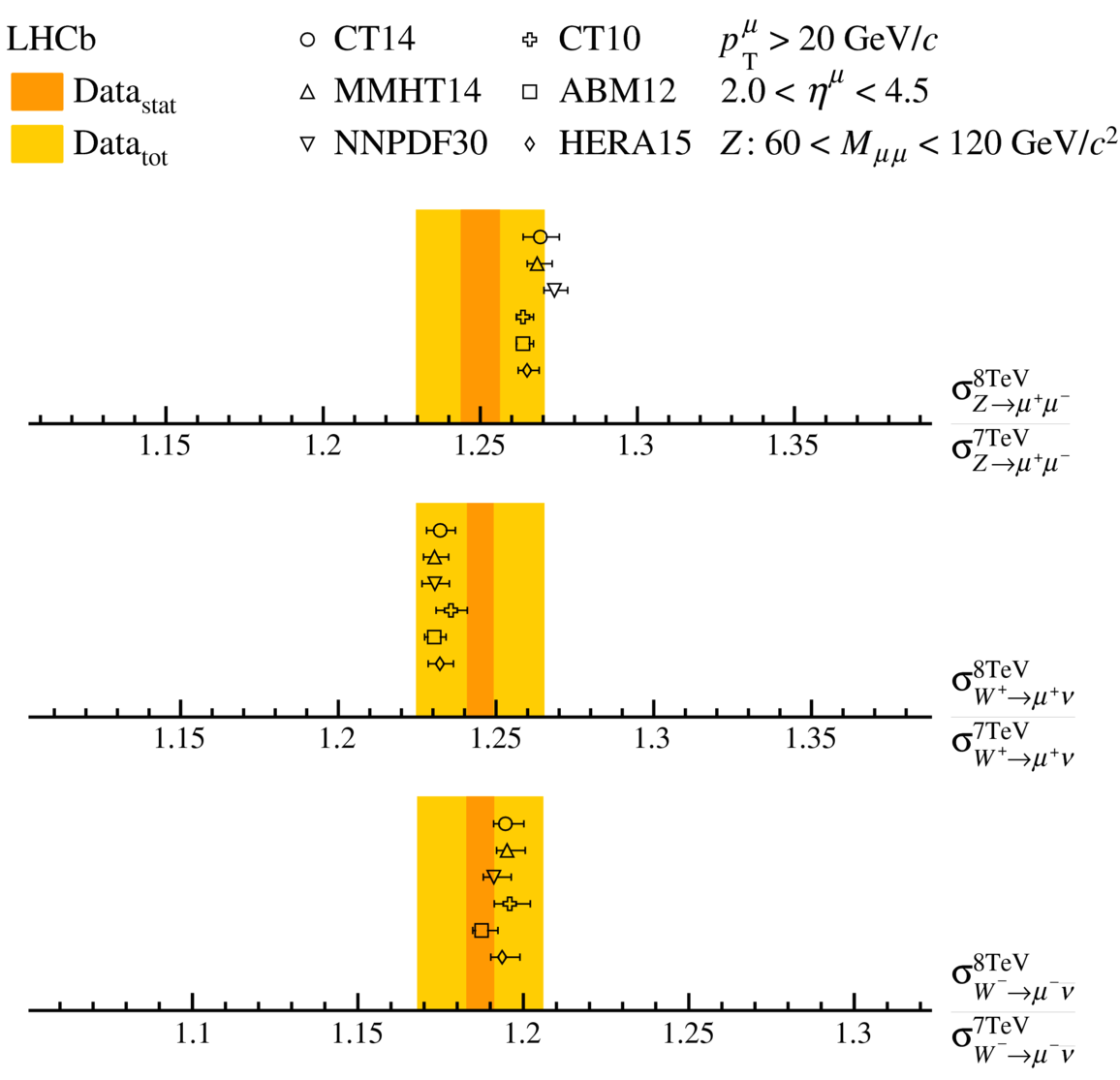}
\caption{Measurements of ratios of different single gauge boson production cross-sections measured by LHCb between 7 and 8~TeV $pp$ collisions. The inner band shows the statistical uncertainty on the measurement, with the outer band showing the total uncertainty. The points correspond to predictions at NNLO using different PDF sets. From the original article~$^{13}$, where more detail is given about the theoretical predictions compared to the data.\label{fig:lhcb_dr}}
\end{center}
\end{figure}

The LHCb experiment has reported a measurement of inclusive $W$ and $Z$ boson production in $7$ and $8$~TeV $pp$ collisions using events containing muon(s) in the final state.\cite{lhcb7,lhcb8} The muons are required to have $2.0<\eta^\mu<4.5$ and $p_{\rm{T}}^\mu > 20$ GeV, and for $Z$ boson production, the dimuon invariant mass is required to be between 60 and 120~GeV. The measured fiducial and differential cross-sections at 7 and 8~TeV, and the charge asymmetries, all show good agreement between NNLO predictions generated using FEWZ~\cite{fewz} and data. The LHCb collaboration has also measured ratios between fiducial cross-section in $pp$ collisions at 8 and 7 TeV (for $W^+$, $W^-$, and $Z$ boson production). These measurements are shown in Figure~\ref{fig:lhcb_dr}. These results show good agreement with NNLO predictions, though a $\sim2\,\sigma$ discrepancy with respect to theory is apparent when ratios of the measurements are evaluated. Future measurements at new collision energies will remain interesting.

\section{Angular distributions of $Z$ boson decay}
We can also consider the decay of directly produced $Z$ bosons at the LHC to a dilepton final state. These measurements are typically performed in the Collins-Soper frame of reference~\cite{cs}, where the differential cross-section may be expressed as
\begin{align*}
  \begin{split}
\frac {\rd^2 \sigma } {\rd\cos\theta^{*}\rd\phi^{*}}
\propto &\Bigl[(1+\cos^2\theta^{*}) +A_0 \frac{1}{2}(1-3\cos^2\theta^{*}) \\&+ A_1\sin(2\theta^{*})\cos\phi^{*} + A_2\frac{1}{2}\sin^2\theta^{*}\cos(2\phi^{*})\\&
  +A_3\sin\theta^{*}\cos\phi^{*} + A_4\cos\theta^{*} + A_5 \sin^2 \theta^* \sin(2\phi^*)\\&+A_6\sin(2\theta^*)\sin{\phi^*} + A_7\sin{\theta^*}\sin{\phi^*}],
\end{split}
\end{align*}
where $\theta^{*}$ and $\phi^{*}$ are the polar and azimuthal angle of the negatively charged lepton. The coefficients $A_0,\;A_1$ and $A_2$ are sensitive to the boson polarisation. At leading order $A_0-A_2 = 0$ (the Lam-Tung relation~\cite{lt}), and deviations from this relation arise from higher-order effects such as multi-gluon emission. The coefficients $A_5 - A_7$ are typically small. $A_3$ and $A_4$ are also sensitive to the vector and axial-vector couplings of the boson. The forward-backward asymmetry in $Z$ boson decays to dilepton finals states, defined by $A_{\rm{FB}}  \equiv \frac{N(cos(\theta^*) > 0) - N(cos(\theta^*) < 0)}{N(cos(\theta^*) >0)+ N(cos(\theta^*) < 0) }$, where $N$ is the number of events fulfilling the relevant condition, is proportional to $A_4$. It can therefore be used to probe the electroweak mixing angle. Such sensitivity is enhanced by the presence of interference from the virtual photon, which generates large asymmetries away from the pole mass of the $Z$ boson. The coefficients $A_5,\;A_6$ and $A_7$ are predicted to be very small.\cite{cmsang}

\subsection{Measurements of angular coefficients}
\begin{figure}
\begin{center}
\includegraphics[width=1.0\linewidth]{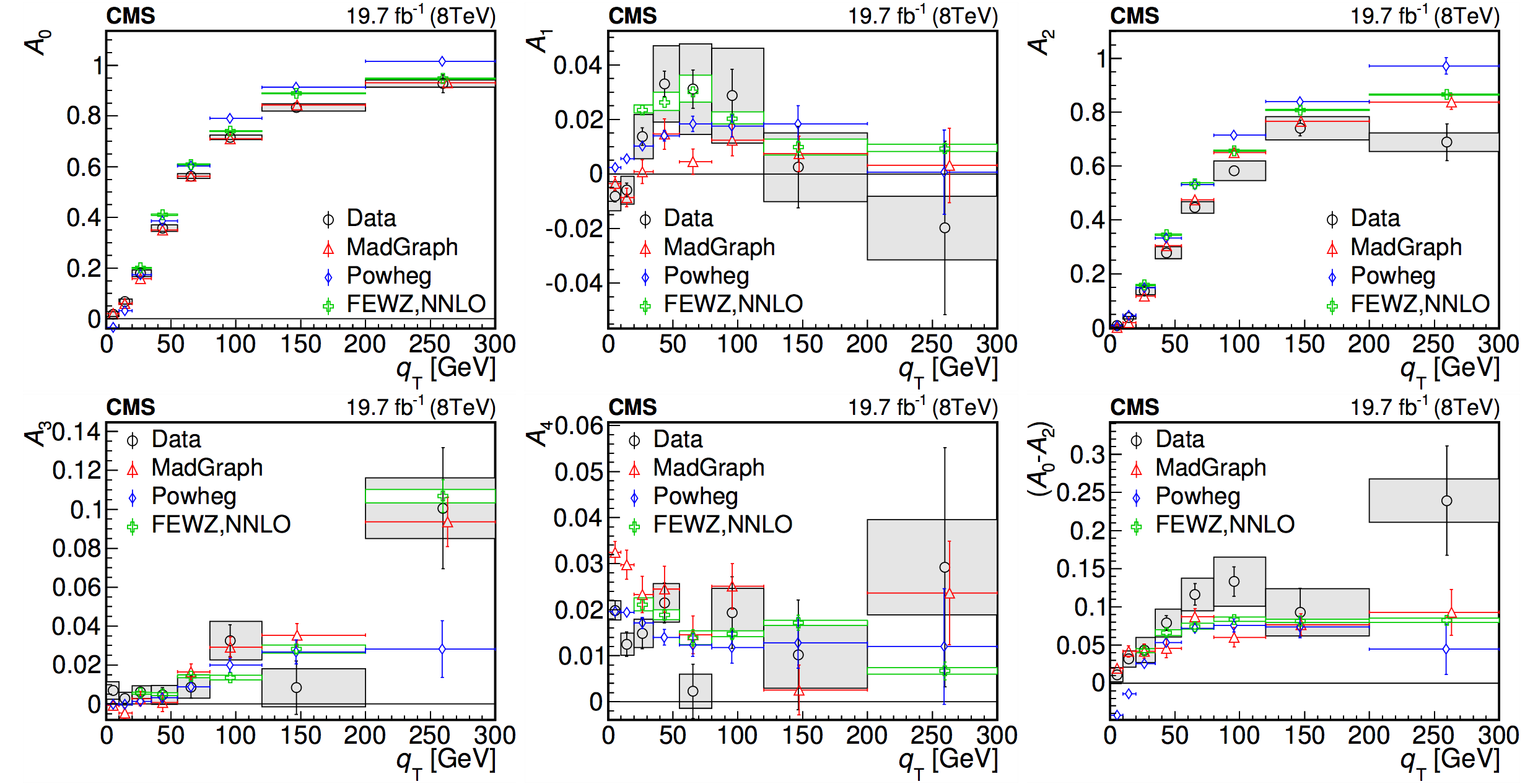}
\caption{Measurements of the angular coefficients as a function of the boson transverse momentum, $q_{\rm{T}}$, for $|y| < 1$. The circles show the measured results, with the vertical error bars showing the statistical uncertainties and the boxes showing the systematic uncertainties. The boxes associated with the FEWZ predictions show the PDF uncertainties. From the original article~$^{16}$, where more detail is given about the theoretical predictions compared to the data.\label{fig:cmsang}}
\end{center}
\end{figure}

The CMS collaboration have reported the measurement of the angular coefficients $A_0$ - $A_4$ in data collected at $\sqrt{s} = 8$~TeV using $Z\rightarrow\mu\mu$ decays, for two regions of rapidity: $|y|<1.0$ and $1.0<|y|<2.1$.\cite{cmsang} The measurements of these coefficients in the central region, as well as $A_0-A_2$, are compared to different theoretical predictions in Figure~\ref{fig:cmsang}. The data do not favour one particular set of predictions, and further refinement of the theory will be needed to better model the data. This will be important for any successful $m_W$ measurement at the LHC.

\subsection{Measurements of the forward-backward asymmetry and determination of $\sin ^2(\theta^{\rm{lept.}}_{\rm{eff.}})$}
\begin{figure}
\begin{center}
\includegraphics[width=0.66\linewidth]{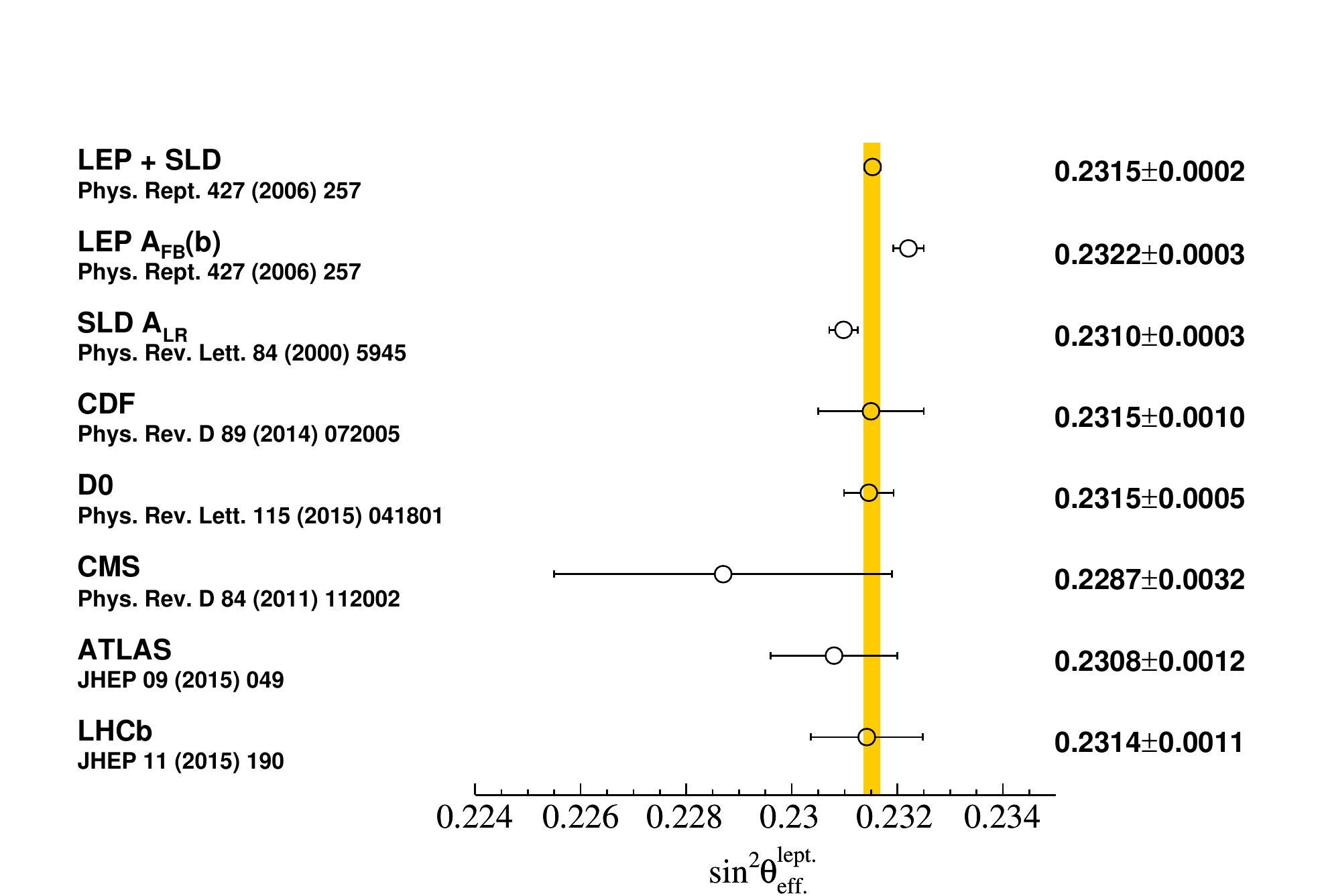}
\caption{Measurements of the electroweak mixing angle at the LHC and at previous colliders. The band shows the combination of the LEP and SLD measurements.$^{20}$ \label{fig:s2tw}}
\end{center}
\vspace{-0.3cm}
\end{figure}

The forward-backward asymmetry in $Z$ boson decays to muon pairs has been measured by the LHCb collaboration in $pp$ collisions at $\sqrt{s} =7$ and 8~TeV, as a function of the invariant mass of the system. This result has then been used to determine $\sin ^2(\theta^{\rm{lept.}}_{\rm{eff.}})$, which is closely related to the electroweak mixing angle, a fundamental parameter within the Standard Model.\cite{lhcbang} The kinematic region probed by LHCb is particularly well suited for this determination, as the phase space has a better defined $z$-axis from which to measure $\cos(\theta^*)$ than the phase space in the central region.\footnote{Systems with boson rapidity $\sim0$ have little sensitivity to $\sin^2(\theta^{\rm{lept.}}_{\rm{eff.}})$ since the direction of the initial state quark has $180^\circ$ ambiguity; the quark direction is diluted in central proton-proton collisions.} The LHCb collaboration reports $\sin ^2(\theta^{\rm{lept.}}_{\rm{eff.}}) = 0.2314\pm0.0007\text{(stat.)}\pm0.0005\text{(syst.)}\pm0.0006\text{(theory)}$. The dominant systematic uncertainty on the extraction of $\sin ^2(\theta^{\rm{lept.}}_{\rm{eff.}})$ is due to knowledge of the detector alignment and momentum scale. This is slightly smaller than the uncertainty in determining $\sin ^2(\theta^{\rm{lept.}}_{\rm{eff.}})$ from $A_{\rm{FB}}$ that arises from PDF knowledge. LHCb is subject to smaller theoretical uncertainties here than the other LHC experiments, since the dilution between the quark direction and the proton direction is smaller in the forward region than in the central region, and is also more precisely predicted with regard to PDF uncertainties; the forward region contains smaller unertainties than the central region from how well any correction for this dilution effect is determined. The result is compared in Figure~\ref{fig:s2tw} to other results from the LHC~\cite{atang,cmstheta} and elsewhere. The two best measurements (from LEP~\cite{LEP} and SLD~\cite{SLD}) are separated by more than $3\,\sigma$. The LHCb measurement is still statistically limited, and future measurements at LHCb and at the other LHC experiments should provide greater precision from improved measurement techniques. Future measurements of $\sin ^2(\theta^{\rm{lept.}}_{\rm{eff.}})$ at the LHC may therefore play an important role in electroweak precision fits.

\section{Conclusions}
Inclusive gauge boson production is a rich field of study. LHC results have placed new constraints on parton distribution functions, and have provided useful inputs for tuning different event generators. Such work is important in its own right, but also paves the way for other crucial measurements of Standard Model physics, such as the measurement of the $W$ boson mass. Studies of the angles at which $Z$ bosons decay to dileptons also allow measurements of boson polarisation and the effective leptonic weak mixing angle.



\section*{References}

\begin{thebibliography}{99}
\bibitem{pdf4lhc}J. Butterworth \emph{et al.}, \emph{PDF4LHC recommendations for LHC Run II}, J. Phys. G: Nucl. Part. Phys. 43 (2016) 023001 
\bibitem{mw}M. D'Alfonso \emph{Prospects for $W$-mass measurements at the LHC}, this conference
\bibitem{mm}M. Mangano and J. Rojo, \emph{Cross Section Ratios between different CM energies at the LHC: opportunities for precision measurements and BSM sensitivity}, JHEP 08 (2012) 010
\bibitem{at13}The ATLAS collaboration, \emph{Measurement of $W$ and $Z$ Boson Production Cross Sections in $pp$ collisions at $\sqrt{s} = 13$~TeV in the ATLAS Detector}, ATLAS-CONF-2015-039
\bibitem{cms13}The CMS collaboration, \emph{Measurement of inclusive $W$ and $Z$ boson production cross sections in $pp$ collisions at $\sqrt{s}=13$~TeV}, CMS-PAS-SMP-15-004
\bibitem{lhcb13}The LHCb collaboration, \emph{Measurement of the $Z\rightarrow\mu\mu$
production cross-section at forward
rapidities in $pp$ collisions at $\sqrt{s} = 13$~TeV}, LHCb-CONF-2016-002
\bibitem{phist}A.~Banfi \emph{et al.}, \emph{Optimisation of variables for studying dilepton transverse momentum distributions at hadron colliders}, EPJC 71 (2011) 1600
\bibitem{at8}The ATLAS collaboration, \emph{Measurement of the transverse momentum and $\phi^{*}_\eta$ distributions of Drell-Yan lepton pairs in proton-proton collisions at $\sqrt{s}=8$~TeV with the ATLAS detector}, submitted to EPJC, arXiV:1512.02192
\bibitem{dynnlo}S. Catani \emph{et al.}, \emph{Vector boson production at hadron colliders: a fully exclusive QCD calculation at NNLO}, PRL 103 (2009) 082001
\bibitem{cms8}The CMS collaboration, \emph{Measurement of the differential cross section and charge asymmetry for inclusive $pp\rightarrow W^\pm X$ production at $\sqrt{s} = 8$~TeV}, submitted to EPJC, arXiV:1603.01803
\bibitem{fewz}Y. Li and F. Petriello, \emph{Combining QCD and electroweak corrections to dilepton production in FEWZ}, PRD 86 (2012) 094034
\bibitem{lhcb7}The LHCb collaboration, \emph{Measurement of the forward $Z$ boson production cross-section in $pp$ collisions at $\sqrt{s} = 7$~TeV}, JHEP 08 (2015) 039
\bibitem{lhcb8}The LHCb collaboration, \emph{Measurement of forward $W$ and $Z$ boson production in $pp$ collisions at $\sqrt{s} = 8$~TeV}, JHEP 01 (2016) 155
\bibitem{cs} J. C. Collins and D. E. Soper, \emph{Angular distribution of dileptons in high-energy hadron collisions}, PRD 16 (1977) 2219
\bibitem{lt} C. S. Lam and W.-K. Tung, \emph{Structure function relations at large transverse momenta in
  Lepton-pair production processes}, PLB 80 (1979) 228
\bibitem{cmsang}The CMS collaboration, \emph{Angular coefficients of $Z$ bosons produced in $pp$ collisions at $\sqrt{s}= 8$~TeV and decaying to $\mu^+\mu^-$ as a function of transverse momentum and rapidity}, PLB 750 (2015) 154
\bibitem{lhcbang}The LHCb collaboration, \emph{Measurement of the forward-backward asymmetry in $Z/\gamma^*\rightarrow\mu^+\mu^-$ decays and determination of the effective weak mixing angle}, JHEP 11 (2015) 190
\bibitem{atang}The ATLAS collaboration, \emph{Measurement of the forward-backward asymmetry of electron and muon pair-production in $pp$ collisions at $\sqrt(s) = 7$~TeV with the ATLAS detector}, JHEP 09 (2015) 049
\bibitem{cmstheta}The CMS collaboration, \emph{Measurement of the weak mixing angle with the Drell-Yan process in proton-proton collisions at the LHC}, PRD 84 (2011) 112002
\bibitem{LEP}The ALEPH collaboration, the DELPHI collaboration, the L3 collaboration, the OPAL collaboration,
the SLD collaboration, the LEP Electroweak Working Group, the SLD Electroweak Group, the SLD
Heavy Flavour Group, the S. Schael et al., \emph{Precision electroweak measurements on the $Z$ resonance}, Phys. Rept. 427 (2006) 257
\bibitem{SLD}The SLD collaboration, \emph{A high-precision measurement of the left-right $Z$ boson cross-section asymmetry}, PRL 84 (2000) 5945
\end{thebibliography}
\end{document}